\documentclass[
reprint,
longbibliography,
superscriptaddress,
preprintnumbers,
nobibnotes,
amsmath,amssymb,
aps,
prl,
floatfix
]{revtex4-1}

\usepackage{gensymb}
\usepackage{mathrsfs}
\usepackage{graphicx}
\usepackage{dcolumn}
\usepackage{bm}
\usepackage{floatrow}
\usepackage{titlesec}

\titleformat*{\section}{\normalsize\bfseries\filcenter}
\titlespacing*{\section}
{0pt}{3.0ex}{2ex}

\usepackage[%
  colorlinks=true,
  urlcolor=blue,
  linkcolor=blue,
  citecolor=blue
]{hyperref}
\usepackage{xcolor}
\usepackage{textcomp}
\usepackage{amsmath}
\usepackage{siunitx}
\usepackage[artemisia]{textgreek}
\usepackage{upgreek}

\usepackage{etoolbox}

\begin{document}
\title{Robust optical readout and characterization of nuclear spin transitions in nitrogen-vacancy ensembles in diamond}

 \author{A.~Jarmola}
  \email{jarmola@berkeley.edu}
    \affiliation{
     Department of Physics, University of California,
     Berkeley, California 94720, USA
     }
      \affiliation{
    U.S. Army Research Laboratory, Adelphi, Maryland 20783, USA 
    }     
     
       \author{I.~Fescenko}
    \affiliation{
    Center for High Technology Materials and Department of Physics and Astronomy,
University of New Mexico, Albuquerque, New Mexico 87106, USA 
    }

\author{V.~M.~Acosta}
    \affiliation{
    Center for High Technology Materials and Department of Physics and Astronomy,
University of New Mexico, Albuquerque, New Mexico 87106, USA 
    }     

 \author{M.~W.~Doherty}
    \affiliation{
  Laser Physics Centre, Research School of Physics, Australian National
University, Canberra 2601, Australia
    }

    \author{F.~K.~Fatemi}
    \affiliation{
    U.S. Army Research Laboratory, Adelphi, Maryland 20783, USA 
    }

     \author{T.~Ivanov}
    \affiliation{
    U.S. Army Research Laboratory, Adelphi, Maryland 20783, USA 
    }

    \author{D.~Budker}
    \affiliation{
     Department of Physics, University of California,
     Berkeley, California 94720, USA
     }
    \affiliation{Helmholtz Institut Mainz, Johannes Gutenberg University, 55128 Mainz,
Germany
    }

 \author{V.~S.~Malinovsky}
    \affiliation{
    U.S. Army Research Laboratory, Adelphi, Maryland 20783, USA 
    }
     
\date{\today}

\begin{abstract}

Nuclear spin ensembles in diamond are promising candidates for quantum sensing applications, including rotation sensing. Here we perform a characterization of the optically detected nuclear-spin transitions associated with the $^{14}$N nuclear spin within diamond nitrogen vacancy (NV) centers. We observe nuclear-spin-dependent fluorescence with the contrast of optically detected $^{14}$N nuclear Rabi oscillations comparable to that of the NV electron spin. Using Ramsey spectroscopy, we investigate the temperature and magnetic-field dependence of the nuclear spin transitions in the 77.5$-$420\,K and 350$-$675\,G range, respectively. The nuclear quadrupole coupling constant $Q$ was found to vary with temperature $T$ yielding $d|Q|/dT=-35.0(2)$\,Hz/K at $T=297$\,K. The temperature and magnetic field dependencies reported here are important for quantum sensing applications such as rotation sensing and potentially for applications in quantum information processing.  

\end{abstract}

\maketitle

Quantum sensors based on nitrogen-vacancy (NV) spin qubits in diamond are used in a number of sensing modalities, including magnetometry, electrometry, and thermometry~\cite{DOH2013, RON2014,DEG2017, BAR2019SEN}. Typically, the qubit used for sensing applications is formed from the NV electron spin levels due to their high sensitivity to environmental perturbations. However, nuclear spins can be more suitable for applications where sensitivity to magnetic noise and temperature variations is undesirable, such as rotation sensing~\cite{LED2012,AJO2012,MAC2012}. Of particular interest are nitrogen nuclear spins intrinsic to NV centers. These spins can be efficiently optically polarized and read out via NV electron spins. 

Consider the example of using the intrinsic $^{14}$N nuclear spins of an ensemble of NV centers for rotation sensing. The $^{14}$N nuclear spins are prepared in a superposition state and precess about their quantization axis with nuclear precession rate $\omega_0$. If the diamond rotates about this axis with a rate $\omega$, the nuclear precession rate in the diamond reference frame is $\omega_0-\omega$. The minimum detectable change in $\omega$ is given by:
\begin{equation}
\begin{aligned}
\label{eq:Sensitivity}
\delta\omega\approx\frac{1}{C\sqrt{\eta NT^{*}_{2}\tau}},
 \end{aligned}
\end{equation} 
where $C$ is the fractional contrast of spin-state-dependent fluorescence, $\eta$ is the photon-collection efficiency, $N$ is the number of interrogated spins, $T^{*}_{2}$ is the spin-coherence time, and $\tau$ is the total integration time. From Eq.~\eqref{eq:Sensitivity}, it can be seen that intrinsic $^{14}$N nuclear spins offer an advantage over electron spins owing to their 10$^3$-fold longer coherence time~\cite{JAS2019} at the same number density. Nuclear spins also have the advantage of having a $10^3\mbox{--}10^4$ times smaller gyromagnetic ratio than electron spins, which minimizes frequency shifts due to fluctuations in magnetic field. 

A remaining challenge is to realize a high spin readout contrast, $C$, without introducing additional sources of technical noise. One avenue that has been explored is to use conditional microwave pulses to map nuclear spin states onto NV electron-spin states~\cite{SME2009, STE2010PRB,NEU2010SCIENCE}. This approach was shown~\cite{JAS2019} to achieve a readout contrast $C\gtrsim10^{-2}$, approaching the contrast realized with NV electron spin ensembles~\cite{BAR2019SEN}. However, environmental influences, such as magnetic field and temperature variations, affect the electron-spin transition frequency, which limits the robustness of this technique~\cite{JAS2019}.

Optical readout of the nuclear spin state can also be accomplished directly without the use of microwave mapping pulses in the vicinity of the excited-state level anticrossing (ESLAC)~\cite{SME2009, STE2010PRB, JAS2019}. The advantage of this technique is that it directly provides information about the nuclear spin states without precise knowledge of the electron spin transition frequencies. While this technique was previously demonstrated, its readout contrast has not been systematically analyzed. 

In this work, we characterize the optical readout mechanism of $^{14}$N nuclear spin ensembles. We find that the contrast of nuclear spin Rabi oscillations exceeds 2\,\% in a broad range of magnetic fields, from approximately 450 to 550\,G. Using Ramsey spectroscopy, we investigate the temperature and magnetic-field dependence of the nuclear spin transitions. At 297 K, we find that the temperature dependence of the nuclear quadrupole coupling constant is $d|Q|/dT=-$35.0(2)\,Hz/K, which is about 2000 times smaller than the temperature dependence of the NV electron-spin zero-field splitting $D$~\cite{ACO2010}. Our results hold promise for quantum sensing applications requiring minimal magnetic field and temperature dependence, including gyroscopes and clocks~\cite{HOD2013}.

 \begin{figure}
\centering
    \includegraphics[width=1.0\columnwidth]{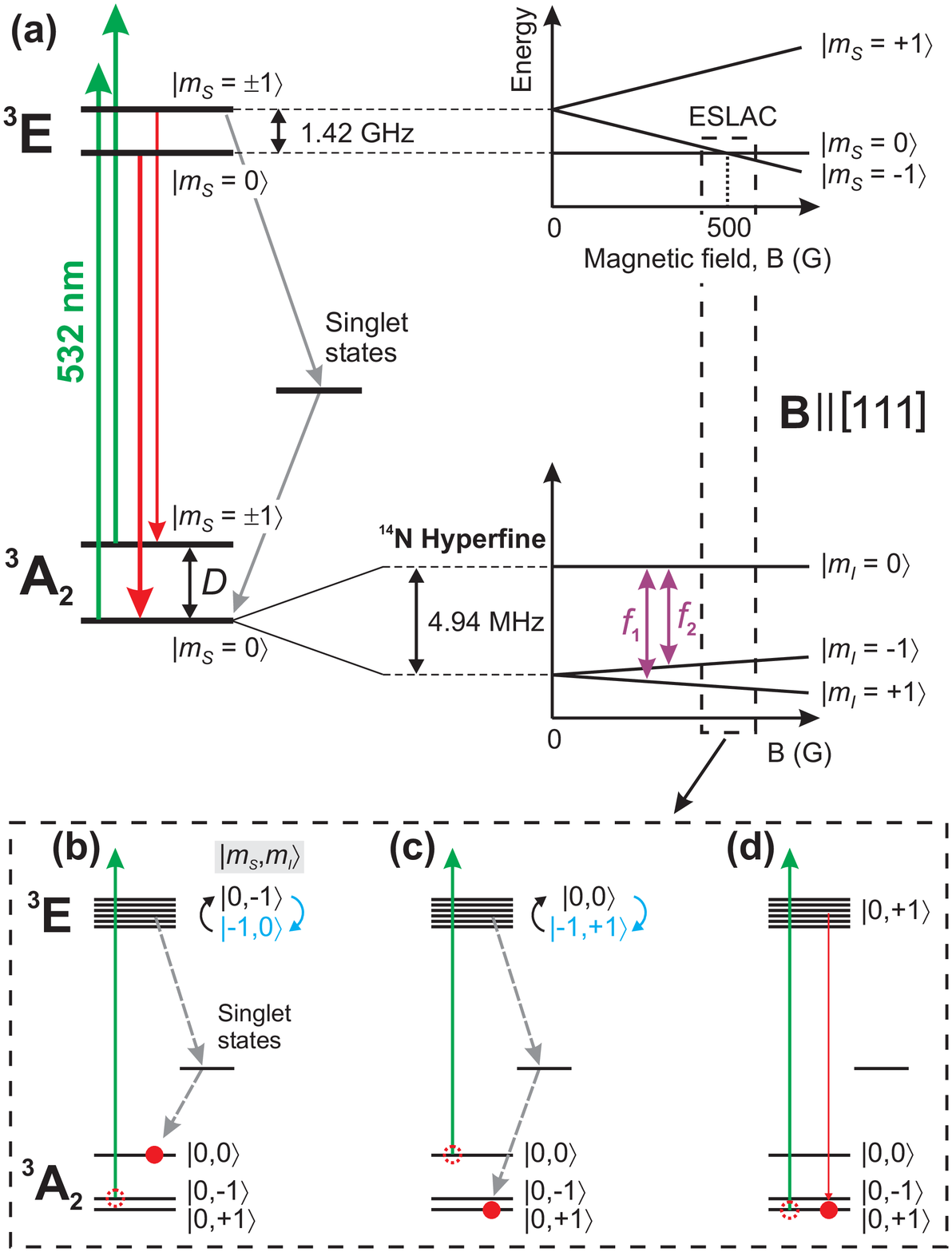}
    \caption{\label{fig:NVlevels} (a) Energy levels of the NV center and its dependence on the magnetic field $\textbf{B}$ applied along the NV symmetry axis. The interaction of the intrinsic electric-field gradient with the quadrupole moment of the $^{14}$N nucleus (nuclear spin $I=1$) leads to a $\sim4.94$\,MHz hyperfine splitting in the $m_{S}=0$ ground-state manifold. In the presence of an applied magnetic field the $m_{I}=\pm1$ nuclear levels split. At $\sim500$\,G the $m_{S}=0$ and $m_{S}=-1$ levels within the excited state undergo a level anticrossing (ESLAC) (b-d) Optical pumping at the ESLAC leads to polarization of $^{14}$N nuclear spin intrinsic to NV center into the $m_{I}=+1$ state. The mechanism responsible for nuclear spin polarization leads to nuclear-spin-dependent fluorescence. 
uorescence}     
\end{figure}

A schematic of the relevant energy levels and transitions in diamond NV centers is presented in Fig.~\ref{fig:NVlevels}(a). Application of light with wavelength shorter than that of the zero-phonon line of the $^3$A$_2\rightarrow ^3$E transition (at 637 nm) induces optical polarization of the NV centers into the $m_S=0$ sublevel of the ground electronic state~\cite{MAN2006}. If a magnetic field $\textbf{B}$ is applied along the axis of the NV center, the $m_S=\pm1$ sublevels of the ground and excited state experience a Zeeman shift. At $B\approx500$\,G the $m_S=0$ and $m_S=-1$ sublevels in the excited state become nearly degenerate. This condition is referred to as excited-state level-anticrossing (ESLAC). A peculiar feature of ESLAC is that in its vicinity, electron polarization is effectively transferred to the nuclei, so that nearly complete $^{14}$N polarization can be achieved in a relatively wide range of magnetic fields~\cite{JAC2009,FIS2013PRB}. A transfer of electron spin polarization to nuclei is also observed at the ground-state level-anticrossing in the vicinity of 1024\,G, but this is not discussed in this work (see Ref.~\cite{AUZ2019} and references therein).

Nuclear spin polarization at the ESLAC mediated by NV centers in diamond has been described, for example, in Refs.~\cite{JAC2009,SME2009,FIS2013PRB,STE2010PRB} and is only briefly summarized here. Near the ESLAC, strong hyperfine coupling in the excited state allows the energy-conserving electron-nuclear-spin flip-flop processes to occur between coupled electron-nuclear spin states, denoted $|m_{S},m_{I}\rangle$. Specifically, such processes can lead to flip-flops between $|0,-1\rangle$ and $|-1,0\rangle$, as well as between $|0,0\rangle$ and $|-1,+1\rangle$ states, Fig.~\ref{fig:NVlevels}(b,c). Under optical illumination near ESLAC the system is polarized into the $|0,+1\rangle$ spin state, Fig.~\ref{fig:NVlevels}(d). 

The mechanism responsible for nuclear spin polarization leads to nuclear-spin-dependent fluorescence and provides the means for direct nuclear spin optical readout. The polarized $|0,+1\rangle$ spin state produces maximum fluorescence, because it is not affected by mixing in the excited state and does not pass thought the dark singlet states as often. As depicted in Figs.~\ref{fig:NVlevels}(b,c) respectively, the $|0,-1\rangle$ and $|0,0\rangle$ states undergo an electron-nuclear-spin flip-flop process in the excited state which changes their electron spin projection to $m_{S}=-1$ and cause them to pass through the dark singlet states, which reduces their fluorescence. The degree of mixing in the excited state, and therefore the fluorescence rate is different for $|0,0\rangle$ and $|0,-1\rangle$ states and depends on the applied magnetic field~\cite{SUP2019}.

\begin{figure*}
\centering
    \includegraphics[width=0.95\textwidth]{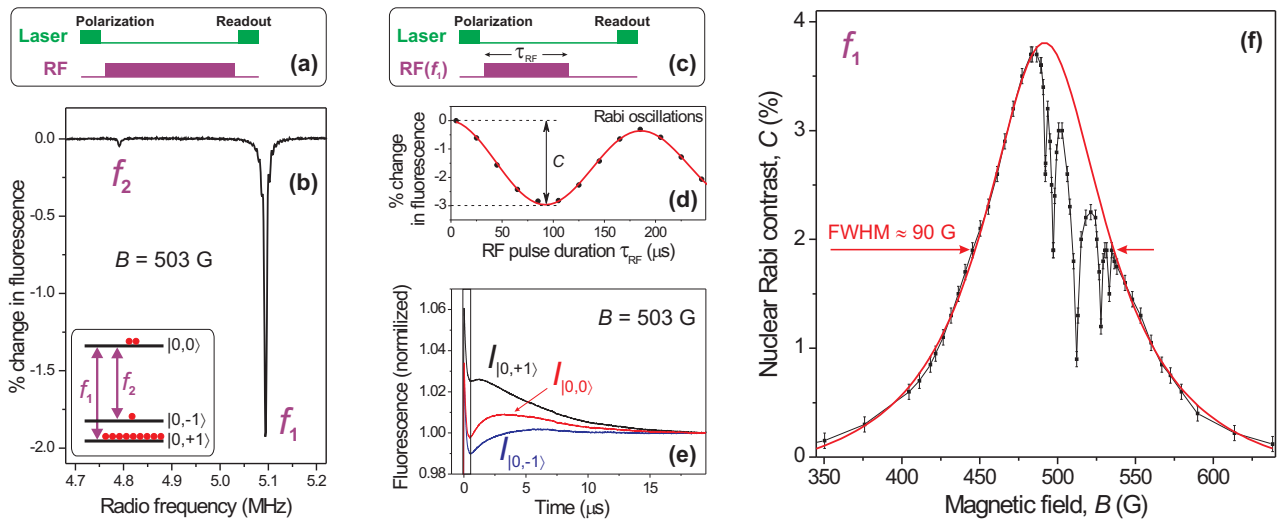}
    \caption{\label{fig:ODNMR} (a) Experimental pulse sequence used for ODNMR spectroscopy. (b) ODNMR spectrum obtained at a magnetic field $B=503$\,G. The inset illustrates the relative population of nuclear levels at this field. (c) Pulse sequence used to observe nuclear Rabi oscillations. (d) Optically detected nuclear Rabi oscillations for the $f_{1}$ transition. Symbols represent experimental data, while solid line is exponentially-decaying sine-wave fit. $C$ - oscillations contrast. (e) Fluorescence responses of the NV centers initialized in $|0,+1\rangle$, $|0,0\rangle$, and $|0,-1\rangle$ states at $B=503$\,G. The vertical lines at the beginning of the traces show the 0.5\,$\mu$s readout time used in the experiments. (f) Contrast of Rabi oscillations between $|0,+1\rangle$ and $|0,0\rangle$ states as a function of the magnetic field $B$. Symbols represent experimental data that are connected with a gray solid line to guide the eye, while the red solid line represents a Lorentzian fit with full width at half maximum (FWHM) of $\sim90$\,G. Five dips observed in the magnetic field range between 490\,G and 535\,G are due to cross relaxation between NV centers aligned along the field direction and substitutional nitrogen spins in diamond~\cite{HAL2016}.    
    }
\end{figure*}

We used a custom-built confocal-microscopy setup to measure optically detected nuclear magnetic resonances (ODNMR) in an ensemble of NV centers. The sample used in our experiments is a [100]-cut high-pressure high-temperature grown diamond with an initial nitrogen concentration of $\sim50$\,ppm. NV centers were created by irradiating the sample with 10\,MeV electrons at a dose of $\sim10^{18}$\,cm$^{-2}$ and subsequent annealing in vacuum at 800\,$^\circ$C for three hours. 

The diamond sample was mounted inside a continuous flow microscopy cryostat. Pulses of 532 nm laser light (20\,mW, 20\,$\upmu$s duration) were focused on the diamond using a microscope objective with 0.6 numerical aperture. Fluorescence was collected through the same objective, passed through a 650-800\,nm bandpass filter, and detected with a fiber-coupled Si avalanche photodiode. Radio-frequency and microwave magnetic fields were delivered using a 100\,$\upmu$m diameter copper wire placed on the diamond surface next to the optical focus. A static magnetic field $B$ was applied along one of the NV axes using a neodymium permanent magnet.

To perform ODNMR spectroscopy we applied a pulse sequence illustrated in Fig.~\ref{fig:ODNMR}(a). The radio-frequency pulse with a typical duration of 200\,$\upmu$s was applied between optical pump and probe pulses and fluorescence response of the system was recorded as a function of the radio frequency. Figure~\ref{fig:ODNMR}(b) shows an example of $^{14}$N ODNMR spectrum recorded at a magnetic field $B=503$\,G. Two resonances were observed at frequencies $f_1$ and $f_2$ that correspond to $|0,+1\rangle \rightarrow |0,0\rangle$ and $|0,-1\rangle \rightarrow |0,0\rangle$ transitions, respectively. The amplitudes of the resonances indicate a strong nuclear polarization of the system into the $|0,+1\rangle$ state.

We used a pulse sequence illustrated in Fig.~\ref{fig:ODNMR}(c) to resonantly drive the nuclear-spin transition with a radio-frequency pulse of varying duration $\tau_{RF}$. Figure~\ref{fig:ODNMR}(d) shows an example of optically detected nuclear Rabi oscillations between $|0,+1\rangle$ and $|0,0\rangle$ states, where $C$ is the contrast of Rabi oscillations.

 We measured the fluorescence responses of the NV centers selectively initialized in the three nuclear-spin states $|0,+1\rangle$, $|0,0\rangle$, and $|0,-1\rangle$ at 503 G, Fig.~\ref{fig:ODNMR}(e). The $|0,+1\rangle$ state was always initialized first by optical pumping, $|0,0\rangle$ state was prepared through transferring the population from $|0,+1\rangle$ state by applying a radio-frequency $\pi$ pulse resonant with $f_1$ transition, and $|0,-1\rangle$ state was prepared by sequentially applying two $\pi$ pulses with radio-frequencies $f_1$ and $f_2$ resulting in transferring the population from optically pumped $|0,+1\rangle$ state.

To get a further insight into optical detection of nuclear spin-states we performed a detailed study of a relative fluorescence responses for $|0,+1\rangle$ and $|0,0\rangle$ states ($f_1$ transition) as a function of applied magnetic field strength near the ESLAC. Figure~\ref{fig:ODNMR}(f) shows the dependence of $C$ for the $f_1$ transition as a function of the applied magnetic field along the NV axis. The contrast of nuclear Rabi oscillations exceeds 2\,\% from approximately 450 to 550\,G, reaching its maximum value of $\sim3.8$\,\% at $\sim 485$\,G.

\begin{figure}
\centering
    \includegraphics[width=1.0\columnwidth]{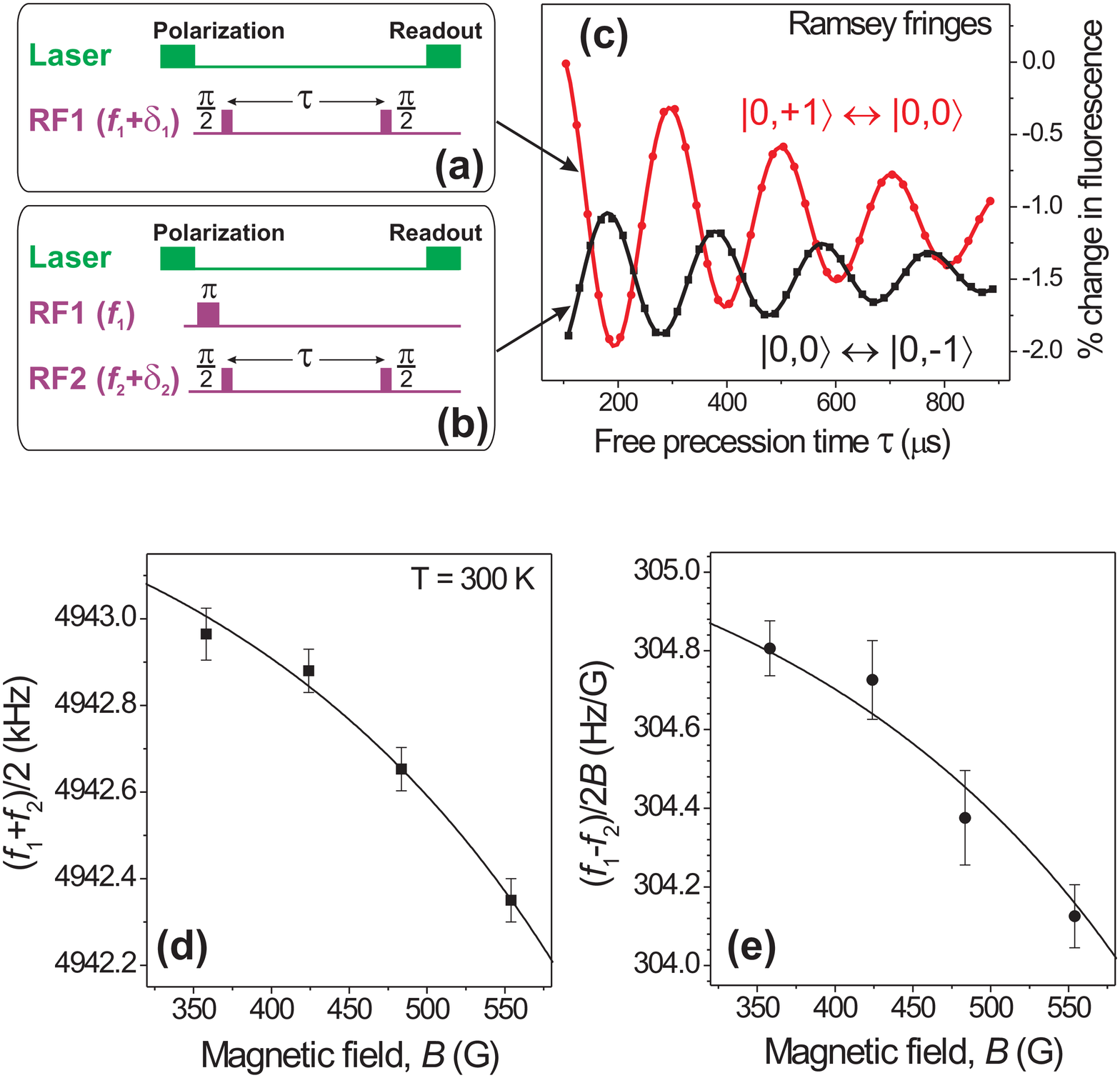}
    \caption{\label{fig:BDep} (a,b) Experimental pulse sequences used to observe nuclear Ramsey fringes for $f_{1}$ and $f_{2}$ transitions, respectively. The radio-frequency pulse sequence $\pi/2 - \tau - \pi/2$ with a variable time delay $\tau$ is applied between laser light pulses. $RF1$ and $RF2$ are detuned from the resonance frequencies $f_{1}$ and $f_{1}$ by $\delta_{1}$ and $\delta_{2}$, respectively. $\pi$ pulse resonant with the $f_1$ transition is used to prepare nuclear spins in $|0,0\rangle$ state. (c) Optically detected NV $^{14}$N nuclear Ramsey fringes for the $f_{1}$ and $f_{2}$ transitions. Symbols represent experimental data, while solid lines are exponentially decaying sine wave fits. The oscillation frequency of the signal corresponds to detuning $\delta$ from the resonance transition $f$, which is $\sim5$\,kHz for the presented data. (d) $(f_{1}+f_{2})/2$ and (e) $(f_{1}-f_{2})/2B$ as a function of $B$. Symbols are values determined from Ramsey spectroscopy, solid lines are fits to Eqs.~(\ref{eq:f1+f2}) and (\ref{eq:f1-f2}), respectively.}
\end{figure}

\begin{figure}
\centering
    \includegraphics[width=1.0\columnwidth]{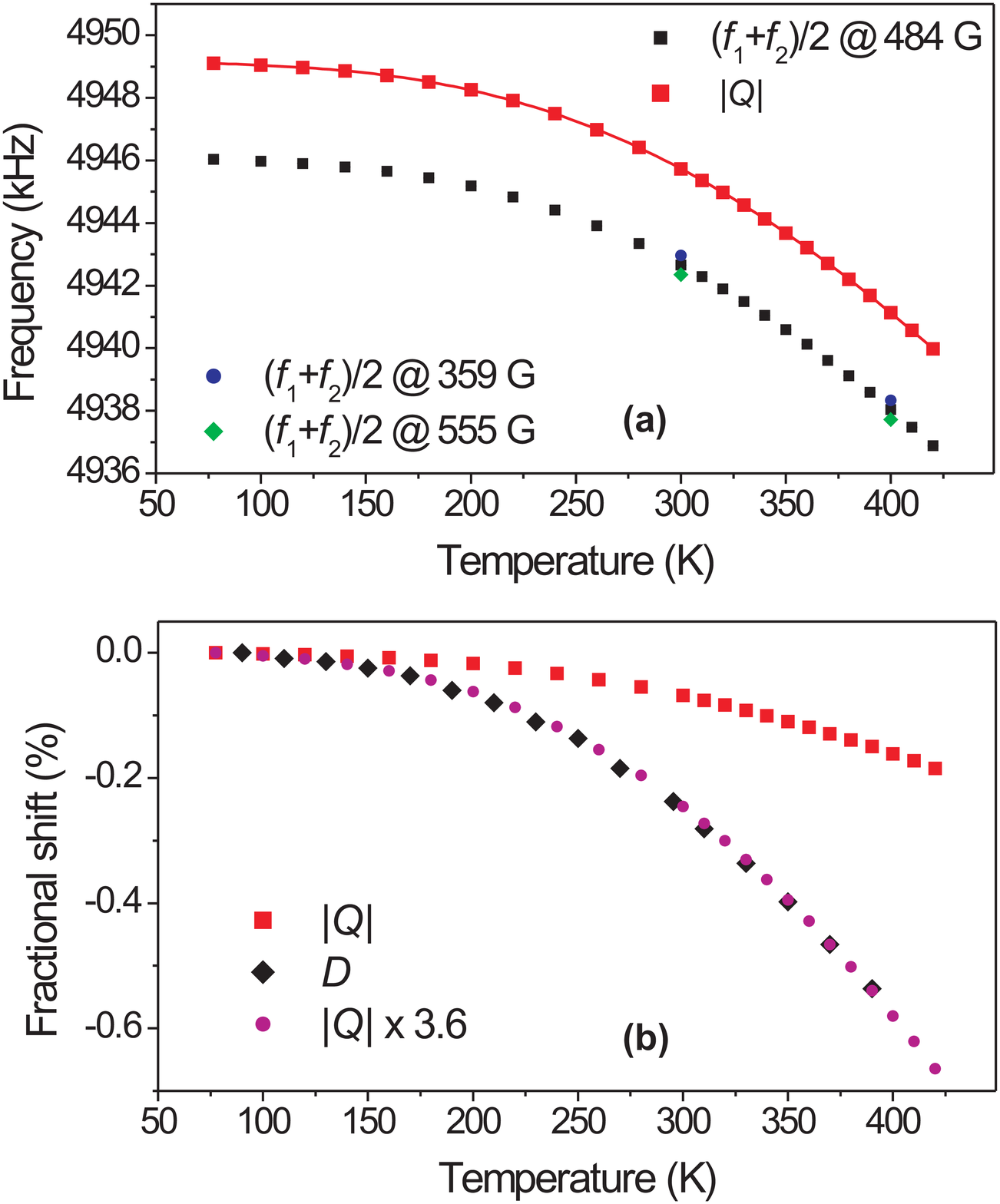}
    \caption{\label{fig:TDep} (a) ($f_{1}+f_{2}$)/2 as a function of temperature at 484\,G magnetic field applied along the NV axis. $|Q|$ was determined from the experimental data using Eq.~(\ref{eq:f1+f2}). Solid line is a forth-degree polynomial fit to Eq.~(\ref{eq:QvsT}). (b) Comparison of the fractional shifts of $|Q|$ and $D$ as a function of temperature~\cite{SUP2019}. }
\end{figure}

Nuclear-spin transition frequencies $f_{1}$ and $f_{2}$ were experimentally measured as a function of magnetic field and temperature by implementing a Ramsey interferometry technique. Figures~\ref{fig:BDep}(a,b) depict the pulse timing diagrams for measuring $f_{1}$ and $f_{2}$, respectively. The $^{14}$N NV nuclear spins are either prepared in $|0,+1\rangle$ state (to measure $f_{1}$) or $|0,0\rangle$ state (for $f_{2}$). Subsequently, a $\pi/2 - \tau - \pi/2$ pulse sequence is applied with the radio frequency tuned near the expected nuclear spin transition. The fluorescence is recorded as a function of $\tau$ and the resulting Ramsey interference fringes Fig.~\ref{fig:BDep}(c) are fit to an exponentially decaying sinusoidal function to reveal the detuning $\delta$ of the transition $f$ with respect to the pulse radio frequency $RF$. From the fit to the Ramsey data, which includes an exponential decay $e^{-\tau/T_{2}^{*}}$, we infer the $^{14}$N nuclear spin-coherence time $T_{2}^{*}$. The experimentally measured values of $T_{2}^{*}$ are in the range from 0.5 to 0.8 ms for all temperature and magnetic field ranges used in this work.

We plot experimental values of $(f_{1}+f_{2})/2$ and $(f_{1}-f_{2})/2B$ as a function of magnetic field in Fig.~\ref{fig:BDep}(d) and (e), respectively. Unlike for an ideal nuclear spin, these values are not constant but, rather decrease with increasing magnetic field strength throughout the studied field range. This is due to mixing of the electron and nuclear spin states via the transverse magnetic hyperfine interaction characterized by the constant $A_{\perp}$. The average value of the nuclear-spin transition frequencies, $f_{1}$ and $f_{2}$ is described by~\cite{SUP2019}:

\begin{equation}
\begin{aligned}
\label{eq:f1+f2}
 \frac{f_{1}+f_{2}}{2}=\left|{Q+\frac{A^{2}_{\perp}D}{D^{2}-\gamma^{2}_{e}B^{2}}}\right|.
 \end{aligned}
\end{equation} 

The effective nuclear gyromagnetic ratio is determined from:

\begin{equation}
\begin{aligned}
\label{eq:f1-f2}
 \frac{f_{1}-f_{2}}{2B}=\gamma_{n}\left (1-\frac{\gamma_{e}}{\gamma_{n}}\frac{A^{2}_{\perp}}{D^{2}-\gamma^{2}_{e}B^{2}}\right ),
 \end{aligned}
\end{equation} 
where $\gamma_{e}$ and $\gamma_{n}$ are the electron and nuclear gyromagnetic ratios, respectively. 

We fit the data plotted in Fig.~\ref{fig:BDep}(d) to the Eq.~(\ref{eq:f1+f2}) with the following parameters fixed: $D=2870$\,MHz, $\gamma_{e}=2.803$\,MHz/G, $A_{\perp}=-2.62$\,MHz~\cite{CHE2015}. 
From the fit we obtained the value of $Q=-4.9457(3)$\,MHz which is in agreement with previously reported values~\cite{SME2009,STE2010PRB} and represents an order of magnitude improvement in precision. We use Eq.~(\ref{eq:f1-f2}) to fit the data in Fig.~\ref{fig:BDep}(e) and extract the $^{14}$N gyromagnetic ratio $\gamma_{n}=307.5(3)$\,Hz/G. The obtained value agrees with the literature data \cite{HAR2002} and confirms the validity of our theoretical model. Error bars of $Q$ represent the combination of statistical uncertainty and the uncertainty in $A_{\perp}$, while in the case of $\gamma_{n}$, the main uncertainty is associated with that in the magnetic field measurement~\cite{SUP2019}.

Next, we measure $f_{1}$ and $f_{2}$ as a function of temperature and use Eq.~(\ref{eq:f1+f2}) and $D(T)$~\cite{SUP2019} to determine $Q(T)$. Figure~\ref{fig:TDep}(a) shows the experimentally measured values of $(f_{1}+f_{2})/2$ and the inferred value of $|Q|$ from Eq.~(\ref{eq:f1+f2}) as a function of temperature. $|Q|$ is shifted from $(f_{1}+f_{2})/2$ by $\sim3$\,kHz and found to smoothly decrease by $\sim10$kHz with temperature increasing from 77.5\,K to 420\,K.

We fit the experimentally determined $|Q|$ data to the forth-order polynomial function:
\begin{equation}
\label{eq:QvsT}
 |Q(T)|=\sum_{n=0}^{4}{a_{n}T^{n}}.
\end{equation} 
The fit values of the coefficients are: $a_{0}=4949.473$\,kHz, $a_{1}=-9.32 \times 10^{-3}$\,kHz/K, $a_{2}=9.2597 \times 10^{-5}$\,kHz/K$^{2}$, $a_{3}=-4.6294 \times 10^{-7}$\.kHz/K$^{3}$, $a_{4}=3.983 \times 10^{-10}$\,kHz/K$^{4}$.
The temperature slope of the nuclear quadrupole coupling constant $Q$ at 297 K is $d|Q|/dT=-35.0(2)$\,Hz/K, which is $\sim2000$ times smaller than the temperature dependence of the zero-field splitting parameter $D$ of electron spin transitions~\cite{ACO2010}, which is $dD/dT=-74.2(7)$ kHz/K. A recent preprint~\cite{SOS2018} used a different technique to infer a value of $d|Q|/dT=-24(4)$\,Hz/K that is lower than our result, but with an order of magnitude larger uncertainty.

 We find that the fractional changes of $Q$ and $D$ as a function of temperature match almost perfectly up to a constant factor of 3.6 (see Fig.~\ref{fig:TDep}b). Thus, the qualitatively identical temperature variations of $Q$ and $D$ suggest that they may arise from a common mechanism. Such a mechanism that would cause similar changes in $Q$ and $D$ is not obvious because $Q$ and $D$ arise from different interactions that depend on different aspects of the NV center's electron orbitals. $D$ arises from the magnetic dipolar interactions between the NV center's two unpaired electrons occupying its $e_{x}$ and $e_{y}$ molecular orbitals, which are exclusively composed of carbon atomic orbitals. Whilst $Q$ arises from the interaction of the $^{14}$N electric quadrupole moment with the electric field gradient at the nucleus generated by the electrons occupying the nitrogen's $p_{z}$ atomic orbital directed along the NV center's axis.
 
The equivalent temperature variations of $D$ and $Q$ present an interesting theoretical problem that will potentially reveal new microscopic understanding of the NV center. Accordingly, it should be pursued in the future with ab initio calculations. The discussions presented in~\cite{SUP2019} may serve as an intuition to guide those calculations.

In this work, motivated by the development of diamond-based rotation sensors, we investigated the nonlinear temperature and magnetic field dependence of the $^{14}$N hyperfine spin transitions in an ensemble of diamond NV centers. These measurements were enabled by a direct optical readout technique (without the use of microwave transitions) optimized in this work. The fluorescence contrast of nuclear Rabi oscillations depends on magnetic field and reaches it maximum value $\sim3.8$\,\% at around 485\,G. Such a high contrast is comparable to that of the electron-spin transitions.

From the magnetic field dependence of the frequencies of the nuclear spin transition, we determine the values of the nuclear quadruple coupling constant $Q$, and gyromagnetic ratio $\gamma_{n}$ for NV $^{14}$N, which are in agreement with the published values.

While the measured temperature dependence of the nuclear-spin transition is smaller than the corresponding dependence of the electron-spin transitions in both absolute (by a factor of $\sim2000$) and relative (by a factor of $\sim3.6$) measure, this temperature dependence can still prove problematic for precision sensors. This dependence can be further reduced by re-configuring the measurement to sense the interval between the $m_I=\pm 1 $ levels directly, in analogy with how this is done for electronic states \cite{FAN2013}. 

We also note that $^{15}$N nucleus does not have a quadrupole moment and therefore there is no quadrupole splitting. Relative advantages and disadvantages of using $^{15}$N vs. $^{14}$N centers for gyroscopic applications require a separate consideration. 

The authors are grateful to Chih-Wei Lai and Pauli Kehayias for useful discussions. This work was supported by in part by EU FET-OPEN Flagship Project ASTERIQS (action 820394), and the German Federal Ministry of Education and Research (BMBF) within the Quantumtechnologien program (FKZ 13N14439) and  A. J. acknowledges support from the Army Research Laboratory under Cooperative Agreement No. W911NF-16-2-0008. M. D. acknowledges support from the Australian Research Council (DE170100169).

%


\bibliographystyle{apsrev4-1}

\clearpage

\begin{center}
\textbf{\large Supplemental material: Robust optical readout and characterization of nuclear spin transitions in nitrogen-vacancy ensembles in diamond}
\end{center}
\setcounter{equation}{0}
\setcounter{section}{0}
\setcounter{figure}{0}
\setcounter{table}{0}
\setcounter{page}{1}
\setcounter{equation}{0}
\setcounter{figure}{0}
\setcounter{table}{0}
\setcounter{page}{1}
\makeatletter
\renewcommand{\thetable}{S\arabic{table}}
\renewcommand{\theequation}{S\arabic{equation}}
\renewcommand{\thefigure}{S\arabic{figure}}
\renewcommand{\thesection}{S\Roman{section}}
\renewcommand{\bibnumfmt}[1]{[S#1]}
\renewcommand{\citenumfont}[1]{S#1}

\section{NV ground-state transitions}
\label{sec:SIham}

The relevant spin Hamiltonian of the NV ground state in the presence of an axial magnetic field $B_{z}$ can be written as:

\begin{equation}
\begin{aligned}
\label{eq:Hamiltonian}
 H=D\left (S^{2}_{z}-\frac{1}{3}\textbf{S}^{2}\right )+\gamma_{e}S_{z}B_{z}+Q\left (I^{2}_{z}-\frac{1}{3}\textbf{I}^{2}\right )-\\
 -\gamma_{n}I_{z}B_{z}+A_{\parallel}S_{z}I_{z}+\frac{A_{\perp}}{2}(S_{+}I_{-}+S_{-}I_{+}),
 \end{aligned}
\end{equation} 
where $\textbf{S}$ and $\textbf{I}$ are the dimensionless electron and nuclear spin operators, respectively, $D$ is the zero-field electron spin-spin interaction, $\gamma_{e}$ and $\gamma_{n}$ are the electron and nuclear gyromagnetic ratios, respectively, $Q$ is the nuclear quadrupole coupling constant and $A_{\parallel}$ and $A_{\perp}$ are the axial and transverse magnetic hyperfine constants. Treating the transverse magnetic hyperfine interaction as a perturbation, the nuclear spin Hamiltonian of the $m_{S}=0$ manifold is to second order:

\begin{equation}
\begin{aligned}
\label{eq:Hamiltonian2}
 H_{0}=\left ( Q+\frac{A^{2}_{\perp}D}{D^{2}-\gamma^{2}_{e}B^{2}_{z}}\right )\left (I^{2}_{z}-\frac{2}{3}\right )-\\
 -\gamma_{n}\left (1-\frac{\gamma_{e}}{\gamma_{n}}\frac{A^{2}_{\perp}}{D^{2}-\gamma^{2}_{e}B^{2}_{z}}\right )I_{z}B_{z},
 \end{aligned}
\end{equation} 
from which we can determine the average value of the nuclear-spin transitions $f_{1}$ and $f_{2}$:

\begin{equation}
\begin{aligned}
\label{eq:Q}
 \frac{f_{1}+f_{2}}{2}=\left|{Q+\frac{A^{2}_{\perp}D}{D^{2}-\gamma^{2}_{e}B^{2}_{z}}}\right|
 \end{aligned}
\end{equation} 
and the effective nuclear gyromagnetic ratio $\gamma_{n}^{\mathit{eff}}$:

\begin{equation}
\begin{aligned}
\label{eq:gamma}
\gamma_{n}^{\mathit{eff}}=\frac{f_{1}-f_{2}}{2B_{z}}=\gamma_{n}\left (1-\frac{\gamma_{e}}{\gamma_{n}}\frac{A^{2}_{\perp}}{D^{2}-\gamma^{2}_{e}B^{2}_{z}}\right ).
 \end{aligned}
\end{equation} 

These expressions are valid sufficiently far from the ground-state level-anticrossing (GSLAC), where the term with the resonant denominator is a small correction. We also note in passing that the the vicinity of the GSLAC is an interesting regime to study the temperature and magnetic field dependence of $A_{\perp}$, which will be a subject of future work.

\section{Nuclear-spin-dependent fluorescence}
\label{sec:Traces}

\begin{figure}
\centering
    \includegraphics[width=1.0\columnwidth]{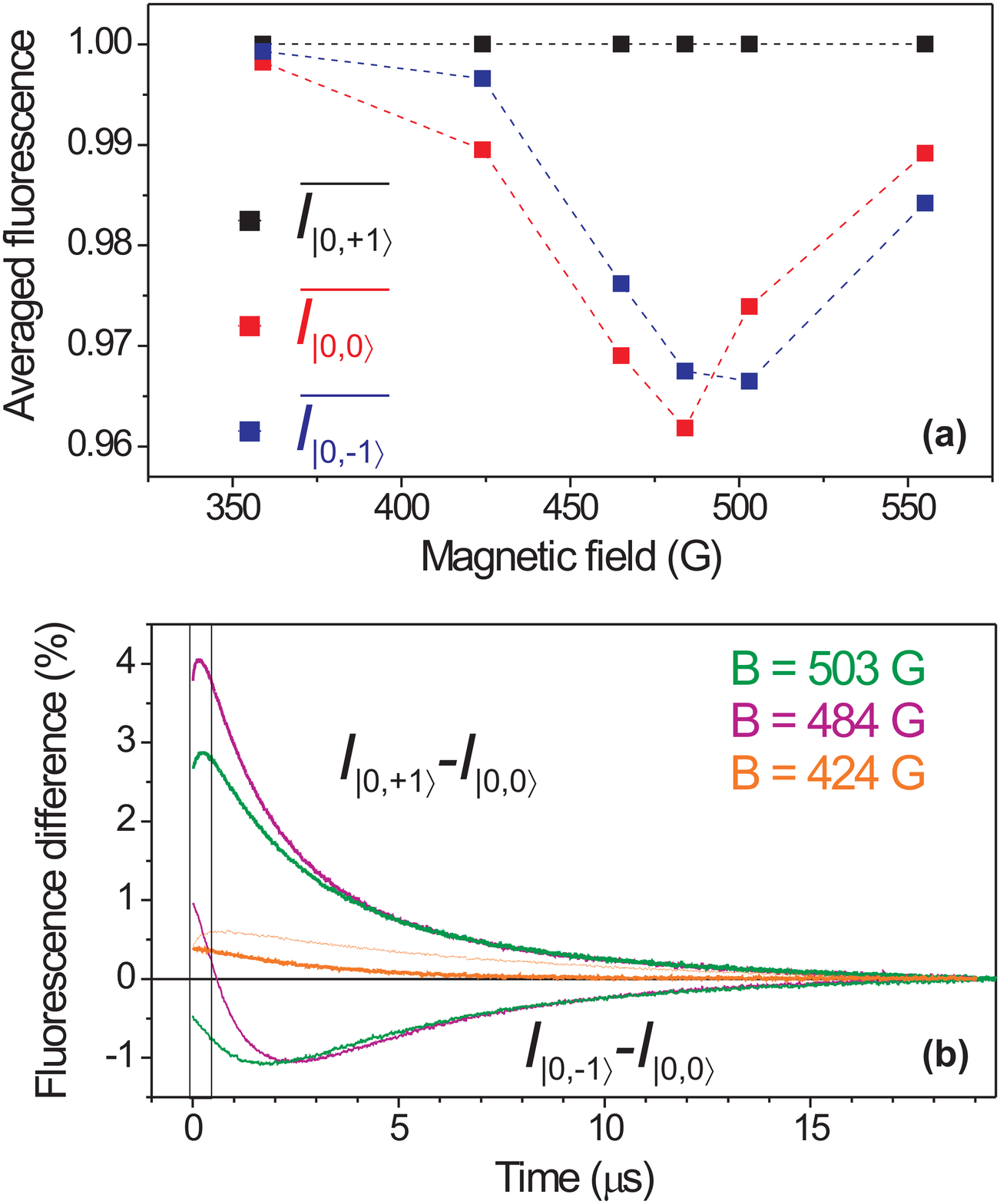}
    \caption{\label{fig:Traces}(a) Normalized averaged nuclear-spin states fluorescence responses (0.5 $\mu$s readout time) as a function of magnetic field. (b) Nuclear-spin states fluorescence response differences at the selected magnetic field strengths (thicker lines: $I_{|0,+1\rangle}-I_{|0,0\rangle}$, thinner lines: $I_{|0,-1\rangle}-I_{|0,0\rangle}$). The vertical lines at the beginning of the traces show the 0.5 $\mu$s readout time used in the experiments presented in the main text.}
\end{figure}

Figure~\ref{fig:Traces}(a) shows an averaged and normalized fluorescence responses (0.5 $\mu$s readout time) of the NV centers selectively initialized in the three nuclear-spin states of the $m_{S}=0$ manifold $|0,+1\rangle$, $|0,0\rangle$, and $|0,-1\rangle$ as a function of magnetic field. The fluorescence rate for the $|0,+1\rangle$ state is the highest for the studied magnetic field range, while the relative fluorescence rate for $|0,0\rangle$ and $|0,-1\rangle$ states depends on the magnetic field strength. The fluorescence rate for $|0,-1\rangle$ state is higher than that for $|0,0\rangle$ state at the magnetic field strength below $\approx495$\,G and it is lower at magnetic field strength above $\approx495$\,G. This relative fluorescence-response behaviour reflects the degree of mixing in the excited state depending on the magnetic field. Figure~\ref{fig:Traces}(b) demonstrates differences in the fluorescence responses for the selectively initialized nuclear-spin states $|0,+1\rangle$, $|0,0\rangle$, and $|0,-1\rangle$ at 503\,G, 484\,G , and 424\,G.

\section{Magnetic field alignment and measurement}
\label{sec:Bmeas}

\begin{figure}
\centering
    \includegraphics[width=1.0\columnwidth]{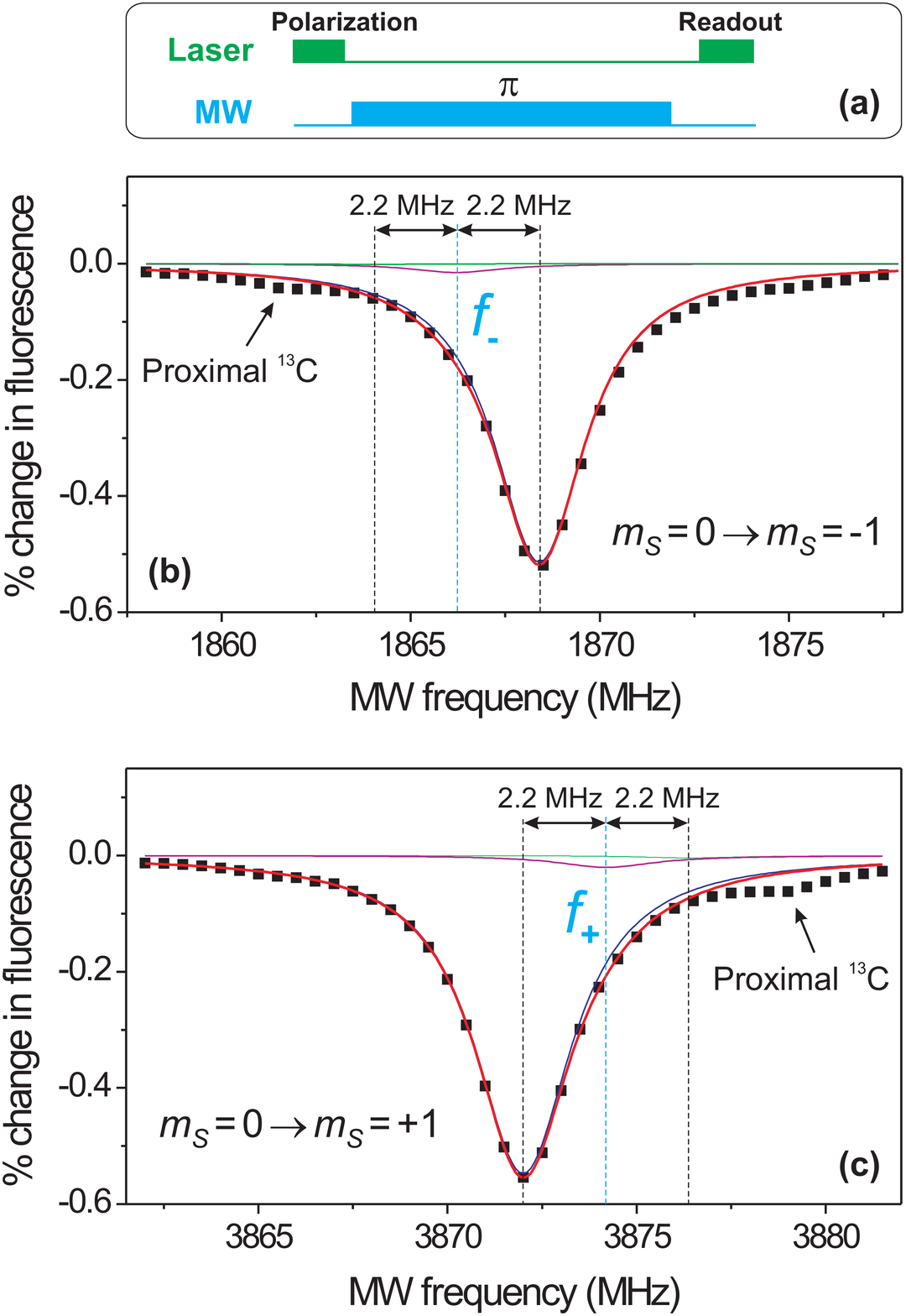}
    \caption{\label{fig:Bcal} (a) Experimental pulse sequence used to obtain ODMR signals. (b,c) Pulsed ODMR signals for the NV subensmbeles aligned along the magnetic field $\textbf{B}$ for $m_{S}=0 \rightarrow m_{S}=-1$ and $m_{S}=0 \rightarrow m_{S}=+1$ transitions, respectively.}
\end{figure}

A static magnetic field, $\textbf{B}$ was applied along the NV axis using a neodymium permanent magnet mounted on a three-axis translation stage. The alignment of the applied magnetic field was done by overlapping three NV electron-spin resonances corresponding to the three NV subensembles which are not aligned with the field. The alignment of the $\textbf{B}$ along the [111] axis is estimated to be better than 0.2 degree. The electron-spin resonances associated with the NV subensemble aligned with $\textbf{B}$ were used to determine the strength of the applied field. Pulsed ODMR signals were recorded for both $m_{S}=0 \rightarrow m_{S}=-1$ and $m_{S}=0 \rightarrow m_{S}=+1$ transitions Fig.~\ref{fig:Bcal}. The duration of the microwave $\pi$ pulse in the pulsed ODMR measurement was on the order of 2\,$\mu$s. Each ODMR signal was fitted with the three Lorentzians separated by 2.2 MHz. The relative amplitudes of the Lorentzians fit indicates the high degree on nuclear spin polarization $>$95\% even at $\approx$150\,G away from the ESLAC. We used the fit values of the central Lorentzian frequency for both transitions $f_{-}$ and $f_{+}$ to determine the strength of the magnetic field:

\begin{equation}
\begin{aligned}
\label{eq:Bcal}
 B=\frac{f_{+}-f_{-}}{2\gamma_{e}},
 \end{aligned}
\end{equation} 
where $\gamma_{e}=2.803$\,MHz/G.
We estimate the statistical uncertainty of the magnetic field measurement to be better than 0.1\,G. Nevertheless, the drift of the magnetic field during the nuclear Ramsey measurements can introduce a systematic uncertainty in $\textbf{B}$. This magnetic field drift associated with the permanent magnet's sensitivity to the ambient temperature fluctuations represents the main source of uncertainty in the measurement of the nuclear gyromagnetic ratio $\gamma_{n}$. We estimate the maximum systematic error introduced by the magnetic field drift to be less than 0.3\,G. 
In the ODMR signal we also observe polarization of the $^{13}$C nuclear spins proximal to the NV centers with hyperfine interaction strength of 13–14\,MHz~\cite{FIS2013PRB}.

\section{Temperature dependence of \textit{D}}
\label{sec:DvsT}

\begin{figure}
\centering
    \includegraphics[width=1.0\columnwidth]{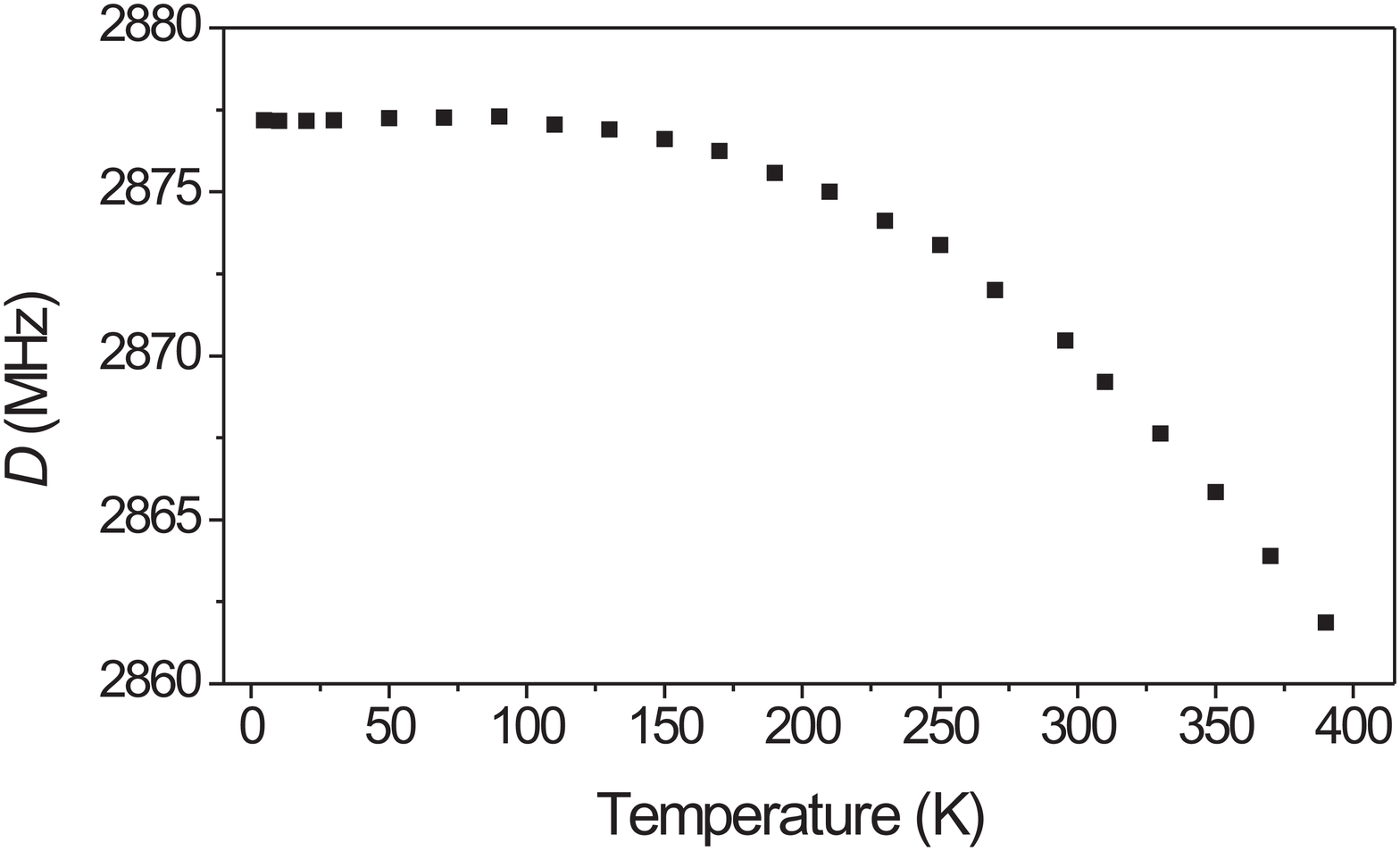}
    \caption{\label{fig:DofT} Experimental values of zero-field splitting $D$ as a function of temperature.}
\end{figure}

Figure~\ref{fig:DofT} shows the experimentally measured values of $D$ as a function of temperature for a diamond sample similar to that used in the present work ([N]$\approx50$\,ppm, [NV]$\approx16$\,ppm) in the range from 5 to 390\,K. These results were obtained employing a standard ODMR technique. The data for the temperature range from 5 to 300\,K were published in~\cite{DOH2014PRB}, while the rest of the data remained unpublished. We use this data to plot the fractional shift of $D$ in Fig.\,4(b) of the main text in the range from 90 to 390\,K. 

\section{\textit{D(T)} vs. \textit{Q(T)}}
\label{sec:DvsT}

To find some link between $D$ and $Q$, we need to also consider how $D$ depends on individual atomic orbitals. As shown in Ref.~\cite{DOH2014PRL}, $D$ is approximately determined by the separation of the mean positions of electrons in the dangling $sp^{3}$ carbon orbitals around the vacancy. These mean positions depend on the carbon lattice positions and the $sp$-hybridization of their dangling orbitals. The lattice positions and hybridization are correlated because if the carbons move relative to their nearest-neighbors, then their atomic orbitals must rehybridize to maintain bonds with those neighbors (this is often called bond bending). The same is true of the hybridization of the nitrogen atom's orbitals and so the occupation of the nitrogen's $p_{z}$ orbital is correlated to the nitrogen's lattice position. Thus, $D$ and $Q$ are linked if there is a common displacement (up to a proportionality factor) of the carbon and nitrogen atoms around the vacancy.

As shown in Ref.~\cite{DOH2014PRB}, the temperature dependence of $D$ has two contributions: thermal expansion and quadratic electron-phonon interactions. These contributions are generalizable across the different types of resonances of the NV center (i.e. visible and infrared)~\cite{DOH2014PRB} and so also expected to govern the temperature dependence of $Q$ (although in that case it will be nuclear-phonon interactions). Both of these contributions represent lattice displacements: either static (thermal expansion) or dynamic (interactions with phonons). Hence, if there are common displacements of the nitrogen and carbon atoms then this may result in the qualitatively identical temperature variations of $D$ an $Q$.

\bibliographystyle{apsrev4-1}


\end{document}